# Theory of pricing as relativistic kinematics

## Melnyk S.I., Tuluzov I.G.

**Abstract**
The algebra of transactions as fundamental measurements is constructed on the basis of the analysis of their properties and represents an expansion of the Boolean algebra. The notion of the generalized economic measurements of the economic "quantity" and "quality" of objects of transactions is introduced. It has been shown that the vector space of economic states constructed on the basis of these measurements is relativistic. The laws of kinematics of economic objects in this space have been analyzed and the stages of constructing the dynamics have been formulated. In particular, the "principle of maximum benefit", which represents an economic analog of the principle of least action in the classical mechanics, and the principle of relativity as the principle of equality of all possible consumer preferences have been formulated. The notion of economic interval between two economic objects invariant to the selection of the vector of consumer preferences has been introduced. Methods of experimental verification of the principle of relativity in the space of economic states have been proposed.

**Contents**

**Introduction**

Currently, an increasing attention is paid to the problems connected with subjective factors of economic relations. Thus, the Nobel Prize in economics in 2014 was awarded to J.M. Tirole for the development of the "Theory of Collective Reputations". The general task of the theory of pricing can be determined as the calculation of a "fair" price for a specific product depending on its quality, volume of transaction, demand for this product and supply of this product. In some currently existing theories the account of other factors influencing the price is possible. In the majority of the existing theories the following assumptions are accepted explicitly or implicitly:

- Fair (true) price for a specific product exists, and it is the same for all consumers (purchasers), thought it can be unknown to them.
- Fair price does not depend on the direction of transaction (purchase or sale).
- The demand and supply are determined unambiguously and do not depend on the price.
- The quality of the product is identical for all purchasers and is characterized not by economic, but by physical parameters, i.e. it is independent.

**The purpose of the present paper** is the construction of the mathematical apparatus, which will allow solving the problem of pricing without these limitations and will be based only on the results of economic measurements, which will be defined further in the paper. Such approach is based on the authors' profound internal conviction that the fundamental approach to the construction of the mathematical apparatus of economics must be based on the properties of symmetry of the space of states of economic objects, which, in turn, are based on the properties of measurements performed on them. Then the equations of dynamics (which are the ultimate objective of any fundamental theory) can be obtained as a sequence of the following scheme (Fig.1). In the present paper we will limit ourselves to the analysis of the first part of this scheme up to the construction of the kinematics of economic objects in the space of states and introduction of the economic invariants. In the final chapter of the present paper we will briefly discuss the perspectives of further development of this theory. Refusal from the aforesaid idealizations in the framework of the discussed (measurement) approach requires not only a formal expansion of the mathematical apparatus, but also a principally new approach to the definition of such notions as equivalence, relative (subjective) price, quality of product, volume of transaction, demand and supply. We will define them regardless of any additional assumptions of the mechanisms of price formation, i.e. only on the basis of the results of economic measurements. Actually, such approach corresponds to the ideology of *geometric dynamics*, which was actively developed in the physical theory in the first half of the 20$^{th}$ century.

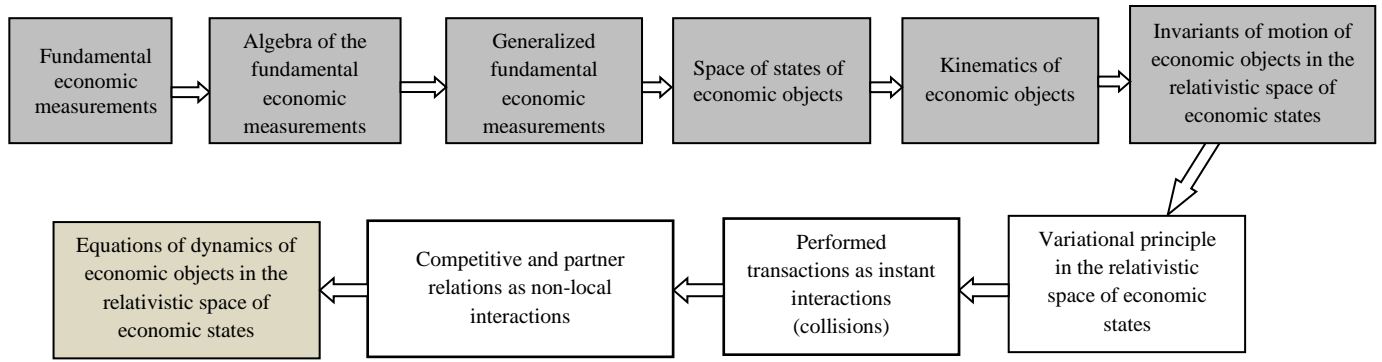

Fig.1. Scheme of construction of the dynamics of states of economic objects on the basis of the measurement approach

### 1. Multidimensionality of the space of economic states

Before proceeding to the description of the theory, let us discuss the principal question on the dimensionality of the economic space. The point is that in the process of its construction we must be guided only by the measurable values. In economics, such value is primarily the price of an economic object. If we express it in conventional units of "ideal money", it will be the only quantitative characteristic of a specific product. On the other hand, products of equal cost can significantly vary in consumer properties. In this case, the necessity of introducing additional (not monetary) parameters of economic objects arises. For the description of the latter, additional dimensions of the space of states are required. We state that these characteristics of "quality" can also be expressed only on the basis of the results of economic measurements (cost of products). However, for this purpose we will have to refuse from illusions of existence of the "true" cost and proceed to a multidimensional space, in which all acceptable estimates of cost will appear to be equivalent.

First, we will illustrate on a simple example the ratio of cost for several economic objects represented in the form of points in one-dimensional and multidimensional economic spaces. The illustration describes to a significant degree the logics of the subsequent steps in the construction of the rigorous theory.

### 1.1. One-dimensional model of the space of states of economic objects based on the measurement approach.

We have previously postulated [1] that the result of a transaction-type measurement is the proportion of exchange of two economic objects. At certain additional idealizations, this definition of the economic measurement allows constructing a trivial space of states of economic objects and calculating the results of transactions.

Let us assume, for instance, that the transaction of exchange of gasoline for sugar is characterized by a certain dimension value $\frac{7}{6}\left(\frac{l.g.}{kg.s.}\right)$. It means that in the conditions of the discussed transaction 7 liters of gasoline (l.g.) are equal to 6 kg of sugar (kg.s.). In this case we can write down the following - $S(7\ l.g.) \equiv S(6\ kg.s.)$, from which it formally follows that $\frac{S(1\ kg.s.)}{S(1\ l.g.)} = \frac{7}{6}$, i.e. the proportion of exchange 7/6 characterizes the ratio of values of 1 kg of sugar and 1 liter of gasoline. Let us also assume that the transaction of exchange of sugar for loafs of bread (l.b.) is characterized by the proportion 3/2, i.e. $\frac{S(1\ kg.s.)}{S(1\ l.b.)} = \frac{2}{3}$. Then we can expect that the transaction of exchange of gasoline for bread will be characterized by the value $\frac{7}{6}\left(\frac{l.g.}{kg.s.}\right) \cdot \frac{3}{2}\left(\frac{kg.s.}{l.b.}\right) = \frac{7}{4}\left(\frac{l.g.}{l.b.}\right)$

Using the properties of logarithmic function and canceling the dimensional units, we can write down: $\log\left[\frac{7}{6}\right] + \log\left[\frac{3}{2}\right] = \log\left[\frac{7}{4}\right]$ and connect each of the summands with the distance between the corresponding economic objects in a certain space (Fig.2). For instance, the summand $\delta l_{gs} = \log\left[\frac{7}{6}\right]$ can be considered as a distance between the economic object "g" – one liter of gasoline and the economic object "s" – one kg of sugar.

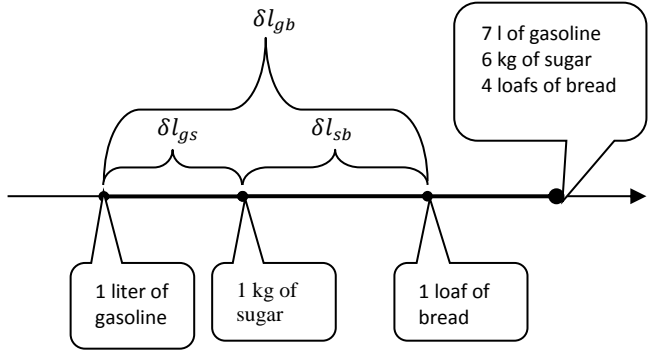

Fig.2. One-dimensional space of economic objects

The we can state that the formula of addition of distances

$$\delta l_{gs} + \delta l_{sb} = \delta l_{gb}$$

fulfils the task set in the theory of pricing – allows calculating the price (proportion of exchange of two economic objects) on the basis of other known prices. Even such a simple ratio allows us obtaining a series on non-trivial results in the process of analysis of an economic system with a preset matrix of technologies [1].

Let us note that in this space we can use any economic object taken in any quantity (for instance, 3 kg of sugar) as a reference point. Then the cost of the remaining objects will be expressed in conventional units of cost equal to the cost of 3 kg of sugar. Thus, the obtained space acts as a uniform price scale. However, it does not represent a number of significant properties of economic objects.

### 1.2. Drawbacks of the one-dimensional model of the space of states of economic objects

The main drawback of the constructed model is the Indistinguishability of economic objects equal in cost (exchanged for each other). Thus, for instance, in the aforesaid example 7 liters of gasoline, 6 kg of sugar and 4 loafs of bread correspond to the same point in the scale of values.

However, in the process of exchange of these objects each of the participants of the transaction assumes that he obtains a bigger value than he returns. Otherwise (in case of equality of these values), the transaction loses its sense. Therefore it is necessary to modify (expand) the one-dimensional space of states in order to fulfill the following requirements:

- Two different exchanged economic objects correspond to different points of the space of states;
- For two participants of the transaction, the object obtained as a result of exchange is to be of bigger value.

### 1.3. Multidimensional space of states of economic objects

These two requirements can be satisfied by introducing a set of various scales of values (one for each of the consumers). Then the ratio "more expensive-less expensive" will depend not only on the position of the objects in the space, but also on the axis (scale), in relation to which they are evaluated.

If for such evaluation we compare the positions of the **projection** of points corresponding to the economic objects, then the possibility of performing transactions, in which each of the consumers considers them profitable for himself, appears. This situation is illustrated in Fig.3

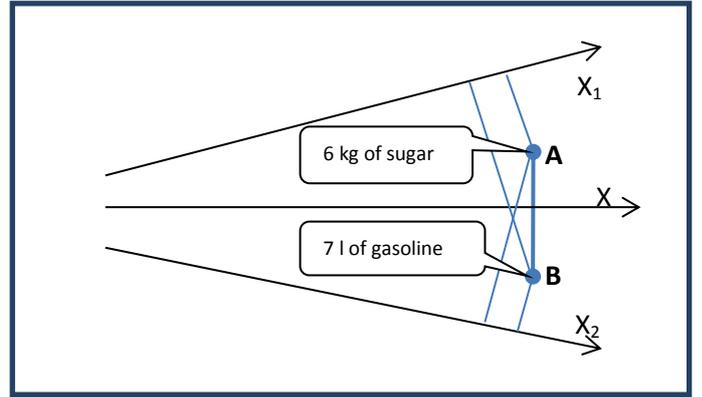

Fig.3 Possibility of a mutually-beneficial exchange in a multidimensional space of economic objects

For consumer X the objects "A" and "B" appear to be of equal value, as illustrated in Fig.1, for consumer $X_1$ «A» is more valuable than «B», but for consumer $X_2$ the situation is reverse.

Any of the consumer directions (vector in the space of states) can be determined by a pair of points. For instance, points «C» and «D» determine the consumer direction «x», for each the project of the segment «CD» (proportion of exchange of these economic objects) is maximum. (Fig.4).

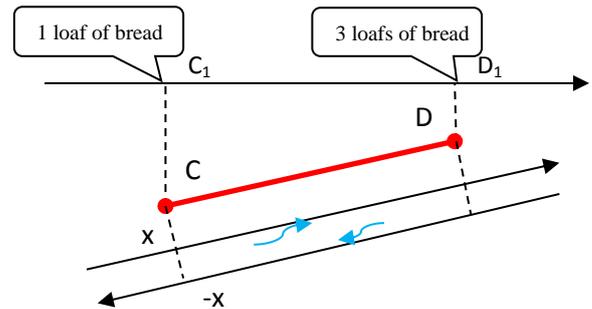

Fig.4. Quantitative estimate of the relative cost of two economic objects («C» and «D» as represented in the Figure) is possible only in relation to the selected scale. Projection of points «C» and «D» on this scale corresponds to the equal objects «$C_1$» and «$D_1$», which differ only in quantity.

For the quantitative definition of the length of the projection $C_1D_1$ it is necessary that points $C_1$ and $D_1$ correspond to the economic objects of equal dimensionality (quantitatively comparable). Therefore, from the whole set of consumer directions we will point out those, which are connected with the economic object (bread, for instance), different quantity of which corresponds to different points of this axis. For instance, point «$C_1$» in Fig. 4 corresponds to 1 loaf of bread, while «$D_1$» corresponds to 3 loafs of bread. Then we can state that according to the "bread" scale of values the economic object "D" is 8 times ($\log_2 8 = 3$) more expansive compared to object "C". Such consumer directions will be further referred to as proper directions.

Let us note that not all possible consumer preferences (directions in the space of state) have an equivalent real economic object, the quantitative scale of which allows measuring the length of projection of the segment. In this case we can state the existence of such consumer direction; however, measurement of the projection on it is impossible.

Further we will propose a consecutive construction of the multidimensional space of economic states in accordance with the scheme illustrated in Fig.1.

## 2. Definition of the main notions of the theory of fundamental economic measurements

Before proceeding to the construction of the axiomatic of fundamental economic measurements, partially developed by us earlier [2-4], let us define the list notions used hereafter. Let us note that they can both match with the notions, generally accepted in the economic theory, and vary from them. Nevertheless, we will further adhere to the definitions given below.

### 2.1. Economic objects

We will define the economic objects as such objects, which can be exchanged for each other (perform transactions with them) completely or partially. These can be both material calculable values (cars, minerals, labor resources) and various services (information, certain actions or refusal from certain actions). Besides, the subject of transactions can be rights and obligations notarized in the form of securities or agreements for non-material assets. In the proposed theory we will not be interested in the physical essence of an economic object or its properties measured in any form other than the economic parameters.

### 2.2. Fundamental economic measurement

We will associate any pair of economic objects with an offer of transaction of their exchange and define it as the fundamental economic measurement. We will denote the fundamental economic measurement, in which a certain subject of economic relations is offered to deliver "B" and receive "A" in return as $[AB]$.

Let us note that the fundamental economic measurements include only the transactions of natural exchange of economic objects. At the same time, in modern economics the majority of transactions are performed indirectly (using money). We will further introduce the additional notion of "ideal money" and analyze their role in the construction of the theory. But first let us discuss only the transactions of natural exchange.

### 2.3. Result of the fundamental economic measurement

We will consider the result of the fundamental economic measurement as a subject's consent for the proposed transaction or refusal of it. At the same time, the result of measurement depends both on the objects of the transaction and on the consumer preferences of subject adopting the decision on the transaction. We will denote the subjects different in their consumer preferences by different small Latin symbols. If a consent is received for the transaction [AB] offered to subject "c", then we will write down the result of this measurement as: $[A_c > B_c]$. In case of refusal, we will write it down as $[A_c \leq B_c]$. Thus, the symbol $A_c$ can be interpreted as an evaluation cost of object "A" by the subject "c". For the transaction [BA] we will obtain $[B_c > A_c]$ and $[B_c \leq A_c]$, accordingly. At the same time, let us note that if $[A_c \leq B_c]$, then $[B_c > A_c]$, however, for different subjects from $[A_c \leq B_c]$ does not follow $[B_d > A_d]$. Moreover, the transaction [AB] can be performed in case if and only if the consent of both subjects of the transaction is received. This means that the simultaneous fulfillment of the two results is required: $[A_c > B_c]$ for the purchaser and $[B_d > A_d]$ for the purchaser.

### 2.4. Indistinguishability of economic objects and fundamental economic measurements

We will consider two economic objects «$A_1$» and «$A_2$» indistinguishable if the result of the transaction $[A_1 B]$ is indistinguishable from the result of the transaction $[A_2 B]$, the result of the transaction $[BA_1]$ is indistinguishable from the result of the transaction $[BA_2]$ for any economic object «B».

We will consider two fundamental economic measurements $[AB]$ and $[CD]$ indistinguishable if for any subject "c" their results will be identical. Either $[C_c > D_c]$ follows from $[A_c > B_c]$ or vice verse, for any "c".

## 3. Algebra of fundamental economic measurements

It is obvious that different economic objects are interconnected by the relation of attribute, as a certain set of economic objects can also be an economic object (can be a subject of transaction). Besides, it is obvious that qualitatively and quantitatively similar (almost indistinguishable) economic objects will almost always have the same results of the fundamental economic measurements.

On the other hand, the results of different fundamental economic measurements can be interconnected by specific conditions and form new transactions with more complex conditions. Essentially, the main task of the theory of pricing is to calculate the results of certain transactions knowing the results of other transactions connected with those transactions in some way.

As the construction of the theory is based on the measurement approach, we will begin the construction of the space of economic states with the construction of the mathematical formalism of interconnections between various fundamental economic measurements, rather than economic objects. We will introduce the binary operations of addition and multiplication of the fundamental economic measurements with obvious (for transactions) properties.

Let us note that the result of operation on two transactions is also a transaction, in which the solution is adopted by one subject. Therefore, the sign of equality in the identities represented below means that the fundamental economic measurements in the right and left parts of the identity give the same results if they will be offered to any of the possible subjects "c".

We will consider the **sum** of two fundamental economic measurements $[A_1 B_1]$ and $[A_2 B_2]$ as a fundamental economic measurement (transaction) $[A_1 B_1] + [A_2 B_2]$, consent for which means that _the consent for at least one of the transactions_ $[A_1 B_1]$ and $[A_2 B_2]$ is obtained. Otherwise (in case of two refusals), we will consider that the transaction $[A_1 B_1] + [A_2 B_2]$ is refused.

It is obvious that in this case the following ratios, which we will accept as axioms defining the properties of the

operation of addition in the fundamental economic measurement, are valid:
- Commutativity $[A_1B_1] + [A_2B_2] = [A_2B_2] + [A_1B_1]$
- Transitivity $([A_1B_1] + [A_2B_2]) + [A_3B_3] = [A_1B_1] + ([A_2B_2] + [A_3B_3])$

If $[0]$ is a transaction which is always refused and $[1]$ is a transaction which is always accepted, then:
- $[A_1B_1] + [0] = [0] + [A_1B_1] = [A_1B_1]$
- $[A_1B_1] + [B_1A_1] = [1]$ or $[A_1B_1] = [1] - [B_1A_1]$
- $[A_1B_1] + [A_1B_1] \neq [A_1B_1]$

We will consider the **product** of two fundamental economic measurement $[A_1B_1]$ and $[A_2B_2]$ as a fundamental economic measurement (transaction) $[A_1B_1] \cdot [A_1B_1]$, consent for which means that _the consent for both transactions_ $[A_1B_1]$ and $[A_2B_2]$ is obtained. Refusal means that at least one of the transactions is refused.

For this operation the following ratios (further referred to as axioms) are valid:
- Commutativity $[A_1B_1] \cdot [A_2B_2] = [A_2B_2] \cdot [A_1B_1]$
- Transitivity $([A_1B_1] \cdot [A_2B_2]) \cdot [A_3B_3] = [A_1B_1] \cdot ([A_2B_2] \cdot [A_3B_3])$
- $[A_1B_1] \cdot [1] = [1] \cdot [A_1B_1] = [A_1B_1]$
- $[A_1B_1] \cdot [B_1A_1] = [0]$
- $[A_1B_1] \cdot [A_1B_1] = [A_1B_1]^2 \neq [A_1B_1]$

Besides, for the pair of the introduced operations the axiom of associativity is valid:
- $([A_1B_1] + [A_2B_2]) \cdot [A_3B_3] = [A_1B_1][A_3B_3] + [A_2B_2] \cdot [A_3B_3]$

### 3.1. Comparison of the algebra of fundamental economic measurements with the Boolean algebra

Let us note that the obtained algebra of the fundamental economic measurements closely resembles the Boolean algebra, as the result of any fundamental economic measurement can possess only two values. At the same time, the significant difference between them is the fact that for the algebra of fundamental economic measurements _the product of two identical transactions_ (indistinguishable in economic sense) means not the same transaction, but a transaction with a doubled volume. If a certain buyer agrees to exchange his property "A" (a bicycle, for instance) for a certain economic object "B" (red telephone), it does not mean that he will agree for a second identical transaction (he may not have a second bicycle or he may not need two identical red telephones).

At the same time, _the sum of two identical transactions_ may not be equal to a single transaction, as it would be in the Boolean algebra. In the aforesaid example it means that the purchaser will agree for two transactions (transaction of doubled volume), but will refuse from each of them separately.

This difference arises because in the Boolean algebra the answer to the same question does not depend on the number of times this question is asked. In the algebra of fundamental economic measurements we have rejected this assumption and we consider that the quantity of positive answers (consents for identical transactions of the same subject) can depend on the quantity of these transactions (volume of the aggregate transaction). Thus, $[A_1B_1]^n$ means a transaction of $n$-times larger volume compared to $[A_1B_1]$, not equivalent to it in the general case.

Let us note that none of the introduced operations allows considering a set of fundamental economic measurements as a vector space, because neither the sum, nor the product of the transactions allow introducing a reverse elements, for which the following identity is valid
$[A_1B_1] + [B_1A_1] = [0]$ or $[A_1B_1] \cdot [B_1A_1] = [1]$

In this connection we will further introduce the notion of the generalized economic measurements, derived from the fundamental economic measurements, allowing to construct the vector space of states of economic objects.

### 4. Generalized economic measurements
#### 4.1. Properties of the scale of volumes of transaction

The continuous scale of volumes of transaction can be obtained in the process of additional studying of the fractional "quantities" of economically indistinguishable transactions. Thus, for instance, we will define the transaction $[AB]^{\frac{1}{2}}$ as a transaction, for which the following equality is valid: $[AB]^{\frac{1}{2}}[AB]^{\frac{1}{2}} = [AB]$. Thus, _any pair_ of economic objects «A» and «B» can be associated with a continuous set of homogeneous transactions.

We will consider the number $\tau = \log_2 n$ as a coordinate on this scale. If we select the transaction $[AB]^k$ as an initial fundamental economic measurement, we will obtain a scale offset by $\tau = \log_2 k$ compared to the first scale. Thus, the selection of the unit of measurement of the volume of transaction is reduced to the selection of a reference point on the logarithmic scale of the "quantity". We will designate the transactions belonging to this set as homogeneous.

The subsets of homogeneous transactions do not intersect, as otherwise we would have a transaction satisfying the condition $[AB]^k = [CD]^l$ in the point of intersection, meaning that $[AB]^{k/l} = [CD]$, and that the transactions $[AB]$ and $[CD]$ belong to the same subset of homogeneous transactions.

#### 4.2. Equivalence of economic objects

Let us consider a pair of economic objects «A» and «B», for which a transaction with positive result (consent of both participants) is possible. This means that at least one consumer "c" exists, who considers that $[A_c > B_c]$, and at least one consumer "d", who considers that $[B_d > A_d]$. Assuming that the consumer preferences are continuously changed, we can state the following. A certain _**set of consumer preferences**_ «s» exists (real or virtually possible consumer), for which these two objects possess equal value $[A_s = B_s]$. The problem of uniqueness of such set will be discussed later.

Let us consider the scale of volume of this transaction. It is obvious (due to the symmetry of the relation of indifference) that for the consumer "s" the objects exchanged for each other in any of the transactions from this scale $[AB]^n$ will be also equivalent. Thus, $[nA_s = nB_s]$.

This means that the considered set of consumer preferences can be associated with an infinite continuous set of pairs of equivalent objects differing only in the parameter $n$. This parameter acts as the volume of the transaction, if we consider the initial pair as the unit volume. The validity of natural axioms is required for the introduced relation of indifference. We will assume that for any subject «s»
- if $[A_s = B_s]$ and $[B_s = C_s]$, then $[A_s = C_s]$
- if $[A_s = B_s]$, then $[B_s = A_s]$

Then it can be shown that the whole set of economic objects relative to any of the subjects of economic relations disintegrates into non-intersecting equivalent subsets. Any pair of these objects allows constructing the scale of volumes of the corresponding transaction. The axioms of the relation of indifference ensure synchronization of these scales.

Let us note that the relations of indifference are associated with the fixed set of consumer preference "s". Therefore, for various subjects the division of the set of economic objects into equivalent subsets and the estimate of the volumes of transaction for them may differ (Fig,5).

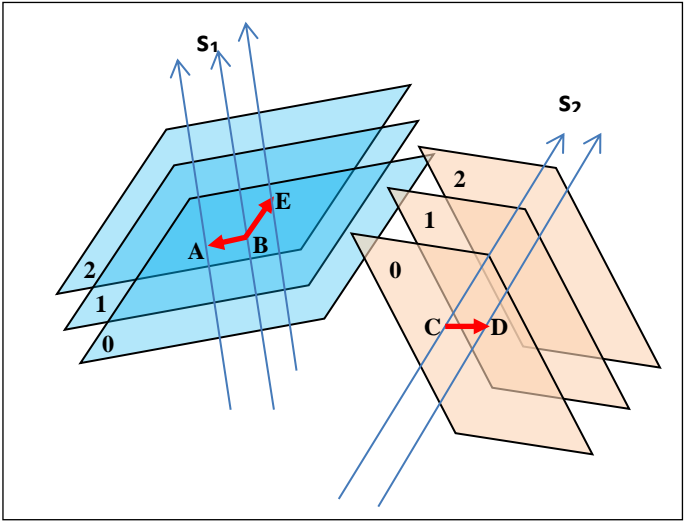

Fig.5. Each set of consumer preference ($s_i$) is associated with a division of the set of economic objects into non-intersecting subsets of objects equivalent for the selected observer. Each pair of the equivalent objects determines the fundamental economic measurement (transaction), which can be used as a basis for constructing the scale of volumes of transactions.

### 4.3. Relativity of estimate of the "quantity" and "quality" of economic objects

It follows from the analysis of the properties of the scale of volume of transactions that all economic objects belonging to one layer (subset of objects equivalent in relation to the consumer "s") are characterized by the same number equal to the volume of the transaction according to the selected scale. For different consumers the same economic object may be associated with different volumes of transactions (Fig.6).

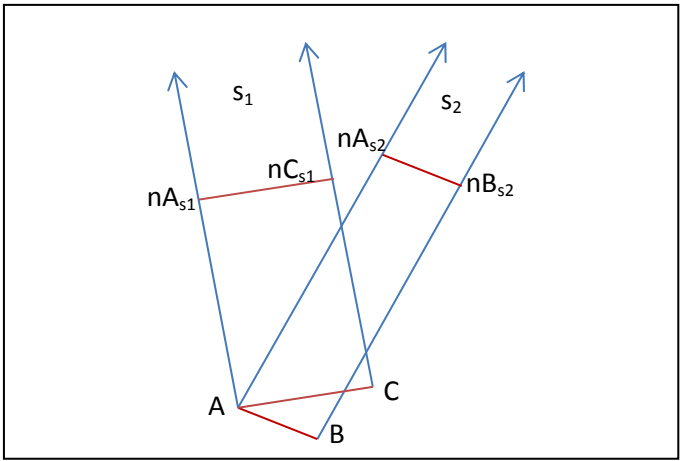

Fig.6. «Quality» of the n-fold «quantity» of the economic object «A» depends on the choice - which of the transactions, $[AB]$ or $[AC]$, is defined as the transaction of unit volume.

Therefore, we will further estimate the volume of a specific economic objects participating in the transaction only in relation to a specific consumer "s" and the corresponding scale of volumes.

However, if from the point of view of a certain observer two economic objects are equivalent (they are associated with the same number on the scale of volume of transactions), then their differences for this consumer can be characterized as "qualitative". We will further illustrate that the availability of the scale of volume of transactions allows estimating these differences quantitatively. At the same time, objects exchanged in the transaction $[A_s B_s]^n$ (we will denote them as $nA_s$ and $nB_s$), from the point of view of the consumer "s" differ from the initial objects $A_s$ and $B_s$ only in terms of "quantity" (volume of transaction), as these transactions form the axes of the scale of quantity (Fig.6).

Thus, any pair of objects, for which an exchange is possible, and the associated transaction [AB] allow to:
- Define the consumer «s», for which these objects are equivalent, thus setting a certain set of consumer preferences;
- Construct the scale of volumes of transactions for this consumer on the basis of the transaction [AB];
- Divide the set of economic objects into non-intersecting subsets of equivalent between each other, but differing in "quality" objects (in relation to the selected consumer);
- Define the subset of identical in "quality" but differing in "quantity" (volume of transaction) objects «$nA_s$» for any object "A" and consumer "s".

Let us note that for different consumers the scale of volumes may appear to be different. Therefore, in the general case $nA_{s1} \neq nA_{s2}$. It means that the n-fold "quantity" of object "A" should be considered as an economically-defined (on the basis of the results of economic measurements for a certain consumer) volume of transaction, rather than its physical quantity. That is why the terms "quantity" and "quality" are used in quotes in this paper.

### 4.4. Proper scale of «quantity»

From all possible scales of volume of transaction, involving the economic object "A", we can mark out such a scale, in which the economic "quantity" of this object matches its "physical" quantity. At the same time, the physical quantity of the object "A" is estimated using a certain internal mechanism not associated with the transactions, for instance, in kilograms or units. We will denote such scale with the index "f". Like other scales, this scale, which we will call "proper scale", can be associated with a certain set of multiple transactions $[A_f B^*]^n = [nA_f nB^*{}_{Af}]$. However, in this case the n-fold quantity of the economic object «B*» - $nB^*{}_{Af}$ appears to be associated with the physical quantity of the economic object "A", rather than with the initial transaction. The advantage of the proper scale in comparison to other scales associated with "A" is that it is defined by the internal properties of the economic object "A", with no connection to other economic objects.

## 5. Space of states of economic objects
### 5.1. Brief analysis of properties of the obtained mathematical structure

The structure of the set of economic objects described above (in chapters 1-3) has been obtained on the basis of the analysis of the properties of fundamental economic measurements and their results for various consumers. It allows performing a certain ordering of the economic objects. However, this ordering is obviously insufficient for solving the main problem of pricing formulated above. Our further aims include:
- Construction of vector space of states of economic objects on the basis of the **generalized economic measurements**,
- Introduction of reference systems associated with specific consumers in this space of states,
- Determining the methods of measurement of the coordinates of "quantity" and "quality" of the product in this space in the selected reference system,
- Deriving of the laws of transformation of coordinates from one reference system into another.

As a result we will obtain the mathematical apparatus for calculating the prices, which are considered fair by specific consumers, allowing us to predict his choice in a particular transaction. The consumer properties in this case can be set by the results of additional measurements. From mathematical point of view, this task is equivalent to the task of calculating the coordinates of a particular point relative to the selected system of coordinates, if its position relative another system of coordinates in the geometric space is known. Actually, we can obtain the geometry of space of economic states. With account of the special character of the axis of "quantity", which in a certain sense acts as the economic time, it can be considered the "economic kinematics".

Rigorous and successive execution of this program first and foremost requires a detailed analysis of the generalized economic measurements. We have previously [2-4] analyzed the logics of transition from fundamental economic measurements to generalized economic measurements. At the same time it has been shown that each such measurement can be correctly described only in the framework of the quantum-mechanical formalism. Let us note that similar results of studying the generalized measurements in physics have been obtained by Schwinger [5]. However, while he postulates their properties, in the process of the discussion of the generalized measurements in economics, the phenomenon of superposition of alternatives occurs as their natural combination in the subject's consciousness in the process of adopting a decision on a transaction. The difference of the superposition of alternatives from their mix occurs due to the fact that in the situation of a delayed choice there is an equal possibility for each of the alternatives of being realized, while in case of their mix we can only speak of a lack of information.

The program of rigorous mathematical construction of the space of states on the basis of the fundamental economic measurements and correct transition to the generalized economic measurements requires a separate research and is beyond the framework of this publication. Our further papers will be dedicated to this problem. Nevertheless, even now we can "guess" the classical limit of such space and verify the conformity of its properties with the natural requirements.

### 5.2. Classical limit of generalized economic measurements

According to N. Bohr, any measurement is based on the comparison with an etalon. Thus, a fundamental measurement can result in one of two answers ("yes" or "no"). Despite this, both physics and other exact sciences operate not with the fundamental measurements, but with secondary measured parameters, such as length, time, etc. We will refer to such measurements as the generalized measurements, meaning their profound interconnection with the theory of generalized measurements of Schwinger [5].

In the application to economics it means that the generalized economic measurements represent a certain combination of the fundamental economic measurements, transformed from them using the previously derived operations of the algebra of measurements. An example of the generalized economic measurements is, for instance, an auction.

Further we will discuss the generalized economic measurements of two types: "ideal sale" and "ideal purchase"

- *«Ideal sale»* assumes that the seller is a monopolist and therefore sales his property at the maximum price, for which the purchasers agree.

- *«Ideal purchase»* assumes that the purchaser is a monopolist and purchases the product at the minimum price offered by the sellers.

At the same time, the price of a certain economic object «$A_0$» relative to an economic object «B» will be considered as the quantity "B", for which it can be exchanged. Thus, in the generalized economic measurement introduced by us, any economic object "A" is associated with the maximum and minimum price according to the scale associated with a set of homogeneous (differing only in quantity) objects "B". These two values of quantity "B" correspond to the generalized economic measurements "Ideal sale" and "Ideal purchase" for the economic object "A".

These generalized measurements combine the results of an infinite set of homogeneous fundamental measurements (transactions of exchange of various quantities of products). The result of these measurements is no longer a consent or a refusal, but a number characterizing the whole set of received answers. Let us note that for obtaining this number it is necessary to construct a scale of quantity of product "B" using a certain set of consumer preferences. Therefore, in the procedure of the generalized economic measurements, a third economic object "C" must be present, which is used for constructing the scale of "quantity" "B" according to the algorithm described above.

As a result it appears that for each pair of economic objects "A" and "B" a certain interval of quantity of economic object "B" exists, for which various consumers are ready to exchange the economic object "A". It is characterized by the limit values: $\tau_{B_{min}}$ and $\tau_{B_{max}}$. This result of the generalized economic measurements can be conventionally written down as an inequality: [$B_{min} \leq A \leq B_{max}$].

Let us note that the limit values no longer depend on the choice of the observer (consumer which evaluates the objects), as in the process of their definition (in the conditions of the transactions of "ideal sale" and "ideal purchase") all possible consumers with various sets of consumer preferences are being questioned. Thus, the obtained characteristics are absolute, unlike the relations of indifference.

Nevertheless, the problem of pricing requires defining not these characteristics, but the subjective relations of indifference of the selected consumer. These relations allow predicting his choice in a particular transaction.

In accordance with this result, we will introduce two numbers characterizing the position of the economic object "A" relative to the scale of the economic object "B" in the space of economic states:

- number $\tau_{A/B} = \frac{\tau_{B_{min}} + \tau_{B_{max}}}{2}$ will be denoted as the volume of transaction («quantity») of the economic object «B», equal to the economic object «A»

- number $l_{A/B} = \frac{\tau_{B_{min}} - \tau_{B_{max}}}{2}$ will be denoted as the distance (difference in "quality") from the economic object "A" to the equal volume of transaction of the economic object "B"

Fig. 7 illustrates the introduced parameters and explains the names selected for them. The quantity of "B" equal to "A" has a transparent economic sense. It is a geometric average (with account of the logarithmic scale) between the maximum and minimum price "A" according to the scale "B".

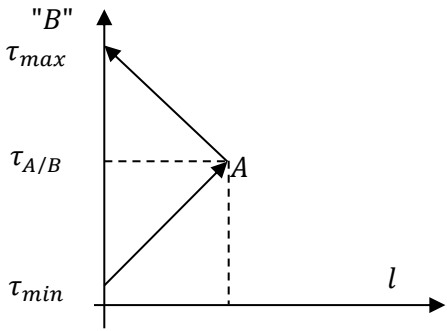

Fig.7. Qualitative difference between the equal quantity of the economic objects "A" and "B" is the larger, the larger is the difference of evaluations of equal quantities of these objects provided by different consumers.

It is clear from the Figure that the larger is the distance from point "A" to axis "B", the larger is the difference between the distance between $\tau_{max}$ and $\tau_{min}$. Therefore, this distance can be determined according to the method described above. On the other hand, following from the general economic considerations, we can conclude that the smaller is the difference of the consumer properties of two products, the smaller will be the difference of prices offered for them by different consumers. Thus, the geometric interpretation of the distance and the economic interpretation of the qualitative differences coincide and correspond to the above mentioned formula.

The factor of principal importance is that the aforesaid definition of equality and distances is self-consistent for three objects, minimally required for performing the generalized economic measurements. Let us first note that if $\tau(B_2)$ is the logarithm of the maximum price (in units B), offered for the object «C», and $\tau(A_4)$ is the logarithm of the maximum price (in units A), offered for the object «$B_2$», then $\tau(A_4)$ is the logarithm of the maximum price which can be received for "C" (Fig.8). Similar statement is valid for minimum prices. Then

$$\tau(C_0) = \tau(B_0) = \frac{\tau(B_1) + \tau(B_2)}{2} =$$
$$\frac{1}{2}\left(\frac{\tau(A_1) + \tau(A_2)}{2} + \frac{\tau(A_3) + \tau(A_4)}{2}\right) = \frac{\tau(A_1) + \tau(A_4)}{2} = \tau(A_0)$$

under the condition that $\tau(A_4) - \tau(A_3) = \tau(A_2) - \tau(A_1)$. However, this condition means that the distance between equal quantities of objects A and B defined as

$$l_{A/B} = \frac{\tau_{B_{min}} - \tau_{B_{max}}}{2},$$

does not depend on these quantities and that the corresponding scales can be constructed on the basis of any of the three pairs of the fundamental economic measurements $[A_0B_0]$, $[A_0C_0]$, $[B_0C_0]$. The obtained axes of "quantity" in the relativistic space of economic states appear to be parallel to each other (as illustrated in Figure 8).

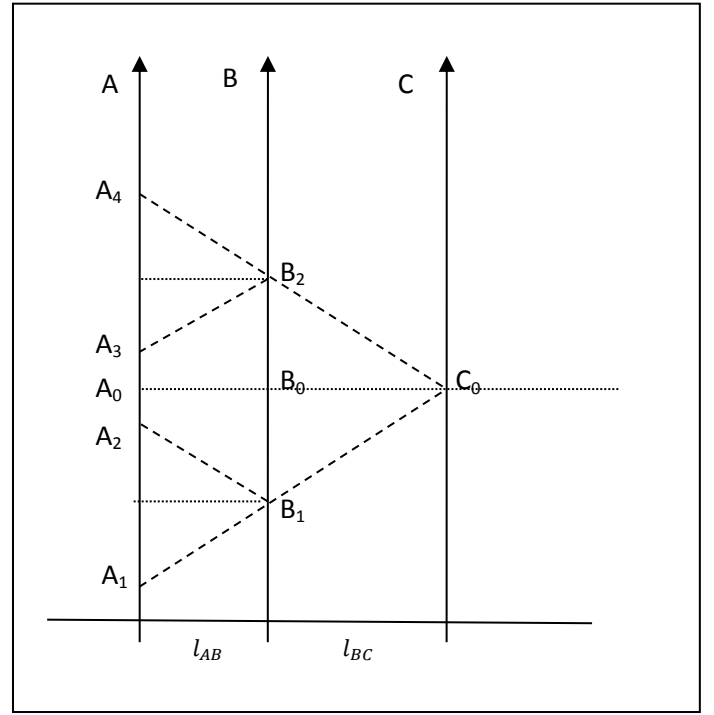

Fig.8. Self-consistence of the definition of equality for different scales corresponding to the same vector of consumer preferences

### 5.3. Vector space of states of economic objects and reference systems in it

Thus, the results of the introduced generalized economic measurements "ideal sale" and "ideal purchase" allow ariphmetize in a consistent way the set of economic objects (set a method of determining their coordinates in the selected reference system). Considering the set of coordinates as a vector and introducing procedures of addition and multiplication by scalar, typical for the vector space, we can construct a vector space of states of economic objects. In the general case, it is multidimensional with a dedicated axis of "quantity" in each of the reference systems.

Comparing the vector space of states with the set of generalized economic measurements, let us note two differences:

- Each transaction (pair of exchanged economic objects) is associated with a certain vector of the space of states, but one and the same vector can be associated with a set of different transactions. Equality of their projections on the axis of coordinates is still insufficient to make them economically indistinguishable.
- The procedure of addition of vectors of the space of states differs from the previously discussed procedure of addition of transactions.

Figure 8 illustrates the two-dimensional space of states of economic objects (one axis of "quantity" and one axis of "quality"). In order to set a reference system in this space, it is sufficient to set a pair of economic objects and a transaction associated with them (the transaction illustrated in the figure is $[A_0B_0]$). Multiple transactions $[A_0B_0]^n$ form the axis of "quantity" or the volume of the transaction, while the set of economic objects equal to each other and to objects $A_0$ and $B_0$ form the axis of "quality". Any of the objects, $A_0$ or $B_0$, can be accepted as a reference point and the distance between these equal (in the selected reference system) objects – as the scale of measurement of distances. Let us also note that the set of reference systems differing only by the reference point and scale correspond to one and the same set of consumer preferences. Therefore, the latter can be considered a vector coinciding with the direction of the scale of "quantity".

In order to complete the construction of the geometric and kinematics in the space of states of economic objects it is primarily necessary to obtain the laws of transformation of coordinates from on reference point into another. For the purposes of reducing the volume of the present publication, we omit the calculations performed by us on the basis of the aforesaid axioms and definitions. Let us note that from mathematical point of view, they are equivalent to the derivation of Lorentz transformations in physics. Therefore, we will further use the obvious analogies with the relativistic physical space and will denote the space of states of economic objects as the relativistic space of economic states.

## 6. Analogies between the physical relativistic space and the relativistic space of economic states

The most reasonable argument for attracting the physical-economic analogies for the purposes of our further analysis is the similarity of the methods of measuring of quantitative and qualitative differences between two economic objects with the measurements of time intervals and distance between two events in physical space. At the same time, the latter can be reformulated in order to avoid the light signal (as it was originally proposed by A. Einstein).

Thus, for instance, if the event "A" is a transmission of a light signal and event "B" is its reception, then we can state that an observer exists, for which the time interval between these events is minimal and tends to 0 (with the increase of the observer's velocity). Thus, the events "A" and "B" for him are almost simultaneous. At the same time, if the event "B" is linked to the state of some physical object (detector), then it is the latest of all events that can be simultaneous with the event "A". And vice verse, the event "A" linked to the source of emission appears to be the earliest of all events, which can be simultaneous with the event "B" for any of the observers. Such formulations completely correspond to the economic definitions of the "ideal purchase" and "ideal sale" given above. Besides, the principle of relativity (absence of a dedicated in space inertial reference system) can be reformulated in the economic context in a practically unchanged form:

❖ *The laws of transformation of quantitative and qualitative characteristics of economic objects (coordinates) must not depend on the consumer preferences of the subject*

Systematizing these and other analogies, let us construct a correspondence table of physical and economic terms in the discussed spaces.

**Table 1.** Correspondence table of the objects of the relativistic physical space and the objects of the relativistic space of economic states

| Physics | Economics |
|---|---|
| Event | Economic object |
| Time interval between events | Differences in the "quantity" of economic objects (logarithm of the ratio of volumes of homogeneous transactions with the participation of the economic objects) |
| Distance between events | Differences in the "quality" of economic objects evaluated according to the width of the range of the proposed proportions of exchange of these economic objects. |
| Simultaneity of two events | Equality of 2 economic objects in relation to a certain consumer |
| Time of a certain event in hours, associated with a certain reference system. | Price of one economic object in the measurement units of another economic object evaluated in accordance with the transaction associated with this economic object. |
| Trajectory of the material point in the selected reference system | Dependence of the coordinate of "quality" of products on the quantity of this product exchanged in one transaction (volume of transaction). |
| Inertial reference frame | Set of consumer preferences (and the associated consumer) |
| Space-like events | Economic objects, which can be equal at least for one of the consumers and cannot be indistinguishable for either of them. |
| Time-like events | Economic objects, which cannot be equal for either consumer, but which are indistinguishable at least for one of them. |
| Proper reference system | Reference system based on the proper (physical) scale of volumes of transaction. |

### 6.1. Non-relativistic limit of the space of economic states

Let us discuss, similarly to physics, the non-relativistic limit of the obtained space and prove that it can be used as an idealized model of the theory of pricing. In physics such transition is performed by means of virtual increasing of the velocity of light up to the infinity. In our model the notion of light signals is not used, but an equivalent effect occurs due to the fact that for different consumers different estimates of value of a particular economic object are possible. These estimates continuously fill a certain interval between $\tau(A_{min})$ and $\tau(A_{max})$. It is obvious that in case of increase of the velocity of light this interval reduces and in the limit turns into 0. The notion of simultaneity becomes absolute and the relativity only influences the selection of the inertial reference system, which is taken as stationary. In this case the law of transformation from one reference system into another is described by the Galilean transformations. Comparison of relativistic and non-relativistic ratios of secondary variable parameters is illustrated in Fig.9.

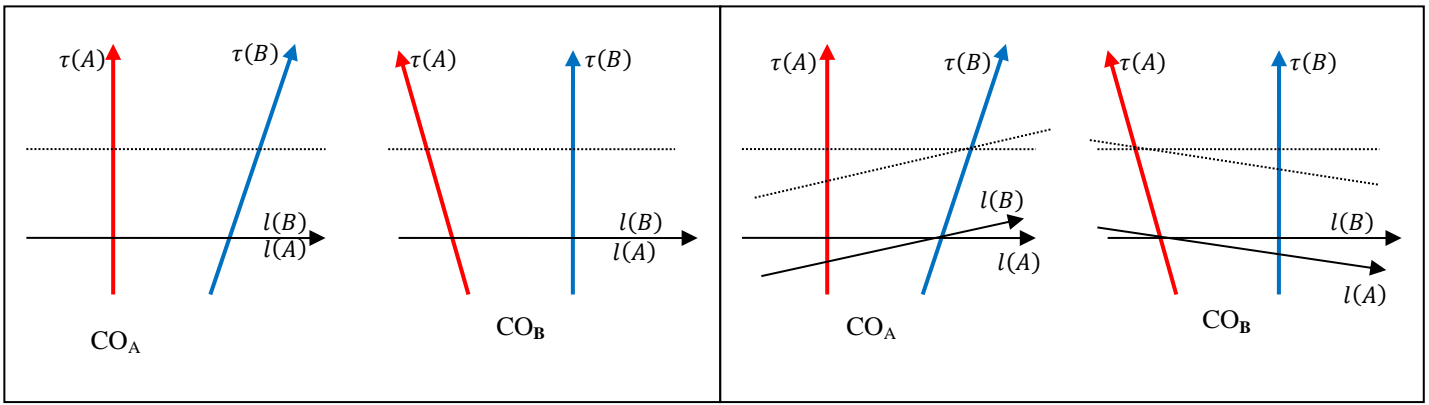

Fig 9. In the non-relativistic space (a) only the space coordinates and velocities are relative. The ratios of equality (analog of physical time) and quantitative differences (analog of physical distance) are absolute. In the non-relativistic space (b) these parameters of motion are relative.

In physics the transition to a non-relativistic limit no longer allows using the light signal for measurement of distances. Instead, the "absolutely rigid line" is used. In the economic space there is no simple analog of such method of measurement. Therefore, it remains unclear how to measure the difference in quality between two equal objects in a non-relativistic limit.

A more substantial drawback of the non-relativistic limit is the fact that in this limit the notion of equality is absolute, i.e. all consumers have absolutely identical opinion of the value of particular economic objects. It completely excludes any profit in transactions and ruins the initial essence of economics (mutual benefit of transactions).

Thus, an adequate consideration of economic relations, unlike physics, is possible only in the relativistic space of states.

### 7. Mathematical apparatus of kinematics in the relativistic space of economic states
#### 7.1. Laws of transformation of coordinates and velocities

For practical use of the kinematics of the relativistic space of economic states it is primarily necessary to derive the laws of transformation of coordinates (results of the generalized measurements), obtained in one reference system, into another. In the economic context it means that if we know the value of the economic object according to the scale of a certain consumer «$S_1$» (associated with the specified set of consumer preferences) and the economic equivalent of its velocity in relation to another consumer «$S_2$», then we can calculate the value of the economic object according to the scale of consumer «$S_2$».

We will not describe the derivation of these ratios as they completely coincide with the laws of transformation of coordinates in the relativistic space (Lorentz transformations). The reason of such coincidence is the mathematical equivalence of the definition of the generalized economic measurements and the mechanism of measurement of space-time coordinates using the light signal in the physical space. Though such coincidence may seem to be "fit" for the already existing formalism, we state that both the initial axioms of economic measurements and the conclusions from them can be obtained on the basis of natural (obvious) properties of the fundamental economic measurements, irrelatively to their physical analogs, only on the basis of the methodology of the information approach.

Therefore, we will further represent these ratios in their standard form (assuming that $c = 1$), but will emphasize on their economic interpretation.

$$\begin{cases} x = \dfrac{x' + vt'}{\sqrt{1-v^2}} \\ t = \dfrac{t' + vx'}{\sqrt{1-v^2}} \end{cases} \quad (1)$$

The coordinates and velocities used in these transformations are secondary results of measurements. Proceeding to the primary results, we obtain, accordingly:

$$\begin{cases} \tau_{max} = \tau_{max}' \sqrt{\dfrac{1+v}{1-v}} \\ \tau_{min} = \tau_{min}' \sqrt{\dfrac{1-v}{1+v}} \end{cases} \quad (2)$$

Let us note that both (1) and (2) are related to measurements in synchronized reference systems with matching reference points. In particular, if $\tau = 0$, then $\tau' = 0$.

At the same time, the notion of velocity remains undefined using the primary results of the generalized economic measurements. It will be discussed in the following chapter. Let us note that in case of changing the direction (sign) of the velocity in the formula (2) $\tau_{max}$ and $\tau_{min}$ interchange symmetrically. It corresponds to the change of the direction of the scale of quantity (analog of time reversion in physics), as could be expected.

#### 7.2. Analog of velocity in the relativistic space of economic states

The principal differences of the relativistic space of economic states from its non-relativistic analog occur as a result of changing of the laws of transformation of velocities. Therefore, the notion of velocity is fundamental, and in this subchapter we are going to discuss its economic meaning. First of all, let us note that both in physics and in economics it makes sense to speak only about relative velocities. They can be expressed using the initial results of measurements. By substituting the expressions for economic analogs of distance and time interval between two events, we obtain:

$$v_{AB} = \frac{dl_{AB}}{d\tau_{AB}} = \frac{d(\tau_{max} - \tau_{min})}{d(\tau_{max} + \tau_{min})} = \frac{k-1}{k+1} \quad (3)$$

where $k = \dfrac{d\tau_{max}}{d\tau_{min}} = \dfrac{1+v}{1-v}$.

Similar formula has been obtained in physics as well [6]:

$$v_{AB} = \frac{k_f^2 - 1}{k_f^2 + 1}$$

If «$A_0$» is the minimum quantity of the economic object "A", for which «$B_0$» can be exchanged, then «$B_0$» is in turn the maximum quantity of the economic object "B", for which «$A_0$» can be exchanged. Making simple calculations, we can obtain the simple ratio for the coefficients $k$: $k_{AB} \cdot k_{BC} = k_{AC}$. And from it, with account of (3), we can obtain the relativistic law of transformation of velocities.

Let us note that the distance (differences in quality) between the object "B" and the equivalent object "A" is completely determined by the ratio of maximum and minimum prices of exchange of "B" for "A" (in the units of measurement of the natural scale "A"). We will call it the relative interval of prices. Therefore, the relative velocity of these objects in the relativistic space of economic states is completely determined by the dependence of the relative interval of prices on the average price. Proceeding from logarithms of prices to the relative prices S, we can obtain the following expression for $k$:

$$k = \frac{d\tau_{max}}{d\tau_{min}} = \frac{dS_{max}}{dS_{min}} \bigg/ \frac{S_{max}}{S_{min}}$$

It follows from this expression that for the relative fixity of two economic objects the following condition must be valid:

$$\frac{S_{max}}{S_{min}} = const \ (S = \sqrt{S_{max}S_{min}})$$

On the contrary, any changing of the relative interval of prices in case of changing the volume of transaction means that the qualitative differences between the objects have changed (depending on the volume of transaction). This is the economic essence of velocity in the relativistic space of economic states. Any variation of this velocity can be associated with a certain force acting on the economic object in the selected reference system

### 7.3. Economic interval in the relativistic space of economic states

Let us note that in the special theory of relativity the statement of the relativity of the results of measurements in different reference systems is non-constructive. It only "prohibits" the absoluteness of the calculable physical parameters (simultaneity of events, length of section, time interval between the events, etc.). At the same time, the events (as a fact that has already happened) remain absolute. In the economic concept developed by us the results of fundamental economic measurements act as such facts. If a certain consumer agrees for a transaction, his consent is absolute and is acknowledged as a fact by all other consumers. However, the evaluation of this fact can be different. Some consumers will suppose that he "made a bad deal".

Therefore, the mathematical apparatus of the theory is based on the requirements of invariance (irrespective of the observer's choice). In physics it is the postulate on the invariance of the velocity of light, and in our theory – the postulate of invariance of the maximum and minimum quantity of the economic object "A", for which a fixed quantity of the economic object "B" can be exchanged.

Formally, we could limit ourselves to these invariant results of economic measurements. However, the obvious image of space and time and the associated trajectory of "motion" arise in the process of the transition to the results of the generalized (calculable) measurements. Therefore, by substituting the expressions of space and time relative coordinates of an economic object into the Lorentz transformations with the help of the absolute results of its measurements, we can obtain an invariant value (interval). Reverse calculations are also possible and allow defining the range of available prices for a particular economic object relative to a random consumer.

Thus, we can calculate the economic analog of the interval between two economic objects "A" and "B" on the basis of the generalized economic measurements using the following formula:

$$(\delta s_{AB})^2 = (\tau_A - \tau_B)^2 - (l_A - l_B)^2 \quad (4)$$

where $\tau_A$, $\tau_A$, $l_A$, $l_B$ are the volume and quality coordinates of the economic objects "A" and "B" relative to any of the consumers. By substituting the expressions of these coordinates using the results of the initial measurements, we obtain:

$$(\delta s_{AB})^2 = (\tau_{A_{min}} - \tau_{B_{min}})(\tau_{A_{max}} - \tau_{B_{max}}) \quad (5)$$

For more obviousness of the obtained formula, we can proceed from the logarithmic scale of volume ($\tau$) to the ordinary scale ($C$). Then we obtain:

$$\delta s_{AB} = \sqrt{\log_2 \frac{C_{A_{min}}}{C_{B_{min}}} \log_2 \frac{C_{A_{max}}}{C_{B_{max}}}} \quad (6)$$

We can also give a verbal definition of the

***Economic interval, as the geometric average of the logarithms of ratios of minimum and maximum prices of two economic objects.***

Its value, unlike the prices, does not depend on the scale, which is used for measuring them.

Let us note that in practice for the estimate of quantity of a particular product its physical scale is normally used, i.e. the volume of transaction is measured in physical values (kg, m$^2$, pieces). Therefore, the consumer associates the scale of values with the physical scale (proper scale) of a particular economic object. As we have previously noted, such scale may not correspond to the inertial reference system, while the Lorentz transformations have been obtained for inertial systems. Therefore, we should expect that the interval calculated according to the formula (3) will remain only for sections of the world line of the economic object that are close to inertial.

We state that in the relativistic space of economic states the value $\delta s_{AB}$ of the economic interval between two economic objects is invariant for various consumers. Thus, in the relativistic space of economic states we had to reject the possibility of finding out the fair (true) price of a particular product and even the fact of existence of such price. Instead, we have obtained a certain invariant, which no longer depends on the consumer preferences and represents an absolute quantitative characteristic of difference of two economic objects, which allows predicting the results of the relative price of these two objects by any of the consumers with the set properties.

### 7.4. Analysis of possibilities of experimental observation of relativistic properties of economic objects in the relativistic space of economic states

First of all, let us note that in physics the majority of relativistic effects are observed only in superaccurate measurements or at large (close to light) velocities. As it has been previously shown, the main criterion of the relative motion of economic objects in the relativistic space of economic states is the dependence of proportions of exchange of two valuable items on the volume of the transaction.

Let us note that in the non-relativistic approximation (Fig.9a) the relation of indifference is absolute. It means that even in case of relative motion of two economic objects their scales of value (proper economic clock) are synchronized and the proportions of their exchange cannot depend on the volume of the transaction. Besides, in this case the maximum and minimum prices should match and the measurement of qualitative differences of two economic objects requires using a classical instrument (analog of a rigorous meter in physics).

Therefore, all economic systems, in which the proportions of exchange depend on the volume of transactions, are relativistic and can be described only with account of the relativistic effects. We should also note that in this case there must be a difference in the estimate by the proprietors of equivalent proportions of exchange of economic objects moving relative to each other.

On the other hand, we should note that both these effects are very weak in the majority of economic systems. Thus, for instance, the dependence of the currency exchange rates on the volumes of transactions is so weak, that it is not normally indicated in the stock exchange quotes. As for the second effect (subjectivity of estimate), such information is mainly confidential and cannot be used directly.

Thus, we can make a conclusion that in real economics, like in Earth mechanics, we often deal with very small velocities of relative motion. Let us assume, for instance, that a twofold increase of the qualitative difference (economic "distance") between two economic objects requires a thousand times increase of the volume of transaction (interval of economic "time"). In this case the relative velocity will be approximately $\frac{\log_2 2}{\log_2 1000} = 0.1$. The relativistic corrections to the expected result (conservation of the proportions of exchange) in this case will not exceed $\left[1 - \sqrt{1 - (0.1)^2} \approx 0.005\right]$.

Therefore, so far we have been unable to find convincing quantitative confirmations of the relativistic character of the relative motion of economic objects in the relativistic space of economic states. Nevertheless, below we will describe some qualitative economic effects, which, as follows from the aforesaid, can occur only if the relativism is present. By calculating the value of the interval between two states of economic objects, we will obtain the possibility of experimental validation of the proposed theory. Let us note that when we deal with proper scales of two economic objects, the formulas 3 and 4 are simplified. Thus, for instance, in case of exchange of two currencies (which can also have different value for different consumers and therefore cannot be considered "ideal money"), we obtain the following relation:

$$(\tau_{B_{max}} - \tau_B)(\tau_{B_{min}} - \tau_B) = (\tau_A - \tau_{A_{max}})(\tau_A - \tau_{A_{min}})$$

With account of the logarithmic scale for the measured prices (visually illustrated in Fig.10), we obtain, accordingly:

$$ln\frac{0.710}{0.705} \cdot ln\frac{0.702}{0.705} \cong ln\frac{1.007}{1.001} \cdot ln\frac{0.996}{1.001}$$

As the maximum and minimum prices differ insignificantly, we can use the linear approximation of the logarithm and obtain:

$$\frac{0.005 \cdot 0.003}{(0.705)^2} \cong \frac{0.006 \cdot 0.005}{(1.001)^2}$$

Like in physics, the sign of the calculated economic interval between the two different economic objects allows unambiguously attribute this pair to the "quality-like" (space-like in the special theory of relativity) or "quantity-like" (time-like in the special theory of relativity).

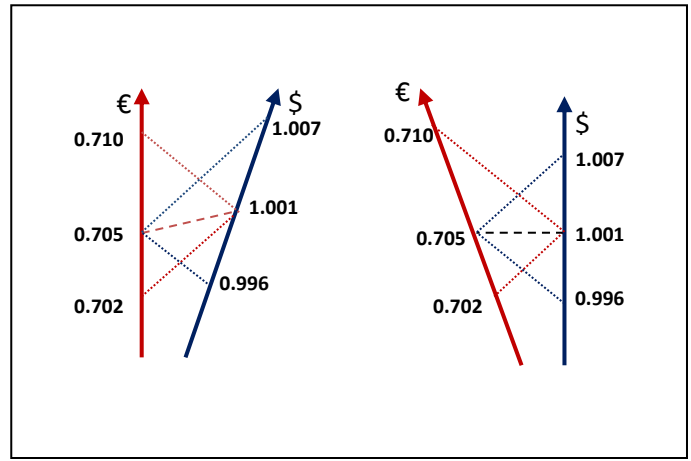

Fig.10. Illustration of the invariance of the economic interval relative to the selection of the inertial reference system.

In the first case we can state that there is at least one set of consumer preferences, in relation to which these objects are equal in value. But there is not a single consumer, in relation to which they are equal in quality.

In the second case, on the contrary, at least one set of consumer preferences exists, in relation to which these objects are equal in quality. But there is not a single consumer, in relation to which they are equal in value.

Let us note that real currency exchange rates very insignificantly depend on the volume of transactions and the maximum price of purchase and the minimum price of sale of a particular currency are measured with large error. Therefore, the values stated above are only an illustration of the principally possible market situation. In real data on currency exchange rates the verification of invariance appear very difficult from the technical aspect. And still, we assume that in the relativistic space of economic states the relativistic effects will occur much more often compared to physics, due to the fact that the differences in the estimate of equivalence (significant relativism of the system) are the incentive reason for concluding mutually-beneficial transactions.

Besides strictly practical applications, the economic invariant introduced by us allows writing down the main equations of "motion" of economic objects in the relativistic space of economic states in the invariant form relative to the consumer, and thus create the basis for constructing the economic dynamics. Further we will briefly describe these stages in accordance with scheme 1. But first we will provide examples of some relativistic economic effects, for the interpretation of which the economic objects must be considered as points in the relativistic space of economic states.

### 7.4.1. Time dilation in a moving reference system (twin paradox) in the relativistic space of economic states

As is known, in case of relative motion of two observers in physics each of them thinks that the clock of the other observer runs behind. This effect is most clearly manifested in case if one of them is moving with acceleration (there and back) relative to the other. Then the clock of the "moving twin" will run behind the clock of the stationary (inertial) observer.

Observation of a similar effect in the relativistic space of economic states requires considering two economic objects, the qualitative differences between which remain identical in certain sections of the scale of "volume of transaction" (Fig.11).

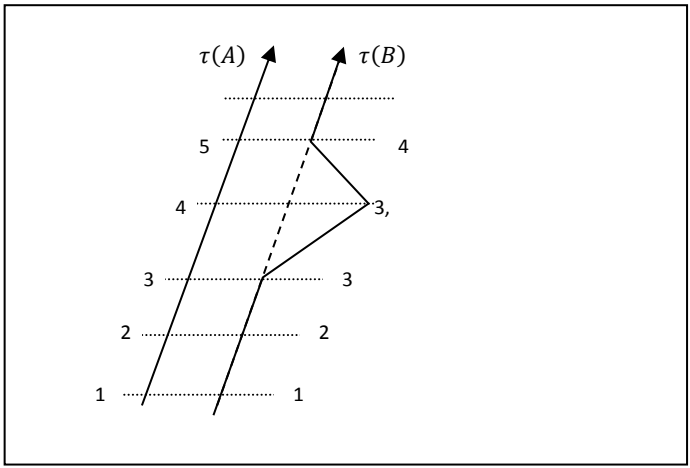

Fig.11. Economic analog of the "twin paradox" shows itself in the changing of proportions of exchange (lagging of the economic "clock") for the economic object, the motion of which is not inertial.

In these sections the "economic clocks" of both economic objects run with the same speed. It means that the increments in the logarithmic scales of the relative price are identical. At the same time, the ratio of the prices (proportion of exchange) does not depend on the volume of the transaction. This condition guarantees the preservation of estimates of equivalent quantities for both proprietors of these economic objects and corresponds to equal velocities of motion of these economic objects in the selected reference system. Besides, the economic distance between equivalent quantities of these objects calculated as the logarithm of relation of the maximum and minimum prices must remain unchanged in case of changing the volume of transaction. We can say that in these sections the economic objects move with identical velocity or that they are fixed relative to each other.

If in a certain section of the scale these proportions at first increase and then decrease (or vice verse) due to the effect of external factors, we can describe it as the relative motion of these economic objects "there and back".

It follows from the Lorentz transformations that after such a section of relative motion the qualitative differences between the "twins" (logarithm of relation of the maximum and minimum prices) return to the initial value and do not change further, however the remaining proportions of exchange become different (economic "clock" of one of the economic objects run behind).

### 7.4.2. Relativistic Doppler effect in the relativistic space of economic states

The difference of the relativistic effect from its classical approximation is that in the relativistic case the relation of frequencies of the transmitted and the received reflected signals is determined only by the relative velocities of the two economic objects and does not depend o their velocity relative to the "environment", like in the classical case. For the experimental verification of this relation we need to perform the economic analog of the Michelson-Morley experiment, which is to demonstrate the absence of velocity of reference system relative to the hypothetical environment – "economic environment". At the same time, like in the original experiment, the main problem is to ensure the constancy of the distance between the "economic mirrors" by independent method. Possibility of performing such experiment and some schemes of its realization in economics will be discussed in a separate publication.

### 8. Possible mechanisms of interconnection of the relativistic space of economic states and physical space-time

So far we have been speaking only about the formal mathematical analogy of the constructed space and the physical relativistic space-time. Various points in the relativistic space of economic states were not associated in any way either with the moment of transaction, or with the distance between the seller and the purchaser. In this connection, a principally different interpretation of the kinematics and dynamics of economic systems in the space constructed by us arises. Thus, for instance, the entire infinite trajectory (world line) of a particular economic object can correspond to one and the same physical moment of time, as it only describes the dependence of the price of transaction on its volume. On the other hand, economic objects located in opposite points of the Earth (for instance, computer programs) can have similar consumer properties and can be located at close points in the relativistic space of economic states.

Altogether, we can state that neither the physical time, nor the physical space is directly connected with the coordinates of economic objects in the relativistic space of economic states. However, a deep metaphysical connection between them exists due to the fact that the fundamental measurements in physics (comparison with the etalon, as Bohr stated) possess properties similar to the properties of transactions in the economic theory. In our opinion, the closest approach to the construction of the physical theory on the basis of the fundamental measurements (binary relations as analogs of generalized transactions) has been made by the followers of the relational approach, for instance, Vladimirov [7]. However, the first and the most fruitful attempt of such construction we can consider the theory of generalized measurements of Schwinger [5].

Nevertheless, in some economic systems the volumes of transactions appear to be connected with the physical time due to technological, rather than economic, properties. As well as the differences according to the scale of economic quality appear to be connected with the physical distance between the objects. Examples, which we are going to represent below, allow visually illustrating the relativistic effects in the discussed economic systems. However, they can be considered as a exception, rather than a rule, and do not allow interpreting the "economic" space-time as a particular case of the physical space-time, and vice verse.

### 8.1. Model of «transportations»

Let us consider a certain economic object – bread, for instance. We will assume that there are several manufacturers of this product, located in different points of the physical space. Moreover, we can consider mobile bakeries as the manufacturers of bread as well. Also, for simplicity, we will assume that all manufacturers produce loafs of bread indistinguishable in their physical properties.

In this situation it is obvious that the purchaser of bread will prefer the product that is manufactured in the vicinity, as its delivery requires less labor and material resources. It is also natural to assume that the cost of additional expenditures for transportation will be proportional to the quantity of the transported bread and the physical distance between the manufacturer and the purchaser. We will write down these assumptions as $S_{B/A} = S_A + k \cdot l_{AB}$, where $S_A$ is the cost (in the conventional units) of one loaf of bread in point $A$, $k$ is the cost of transportation of one loaf of bread for 1 km, $l_{AB}$ is the distance between points A and B, $S_{B/A}$ is the cost of one loaf of bread produced in A for the consumer located in point B.

In order to reduce the sum of overall expenditures, the transporter can spend the resources more optimally. For this purpose he must buy them not in the beginning of the route, but as may be necessary, at the same time exchanging part of the transported bread. Then for a small section of the distance $dx$ we can write down: $S(x)dn = k \cdot n(x) \cdot dx$, where $S(x)dn$ is the funds received for the sale of a part of the amount of bread, $n(x)$ is the quantity of bread remaining after the sale. From this equation it follows that: $\frac{d}{dx}\ln n = \frac{k}{S(x)}$. Let us make one more assumption that all consumers and all points in the map, in which they are located, are equivalent. It means that the cost of a loaf of bread located in the same point with the consumer will be the same for all points. We will denote it as $S_0$. By integrating the aforesaid expression along the whole route, we obtain:

$$\ln \frac{n_B}{n_A} = \frac{k}{S_0} l_{AB} \qquad (7)$$

Thus the consumer located in point B can exchange a certain valuable item $B_0$ for different quantity of bread produced in A depending on the question who is paying for the transportation of bread. If the transportation is paid by the seller (proprietor A), the price of bread is minimum and equals $A_{min}$. If it is paid by the purchaser (proprietor B), it is maximum and equals $A_{max}$. The rest of the variants of agreements of exchange will provide intermediate values. In the "compromise" variant of exchange each of the participants of the transaction pays for his half of the distance of transportation (the exchange is performed in the middle of the distance AB). In this case, the quantity of bread exchanged for $B_0$, equals $A_0 = \sqrt{A_{min} A_{max}} \cong B_0$. At the same time, $\ln A_0 = (\ln A_{min} + \ln A_{max})/2 \cong \ln B_0$. In accordance with the previously introduced definition, the quantity of bread $A_0$ is equivalent to the value of $B_0$ according to the scale associated with the bread produced in $A$. Besides, it follows from (7) that $\ln \frac{A_{max}}{A_{min}} = 2 \frac{k}{S_0} l_{AB}$. Therefore, the value $\frac{(\ln A_{max} - \ln A_{min})}{2} = \frac{k}{S_0} l_{AB}$ can be used as a economic method of measurement of the distance $l_{AB}$. In the discussed model the value $S_0/k$ acts as the "economic velocity of light" and $\ln A_0$ as the "economic moment of time" simultaneous with the "economic event" $B_0$. As $k$ depends on the unit of measurement of the distance, we can select this unit so as to satisfy the equality $S_0/k = 1$. Then the numerical values of the physical and economic distance will be identical.

It is obvious that the value $\ln A_0$ does not depend on the distance, in which a stationary bakery A is located from B. However, if we consider a mobile bakery $A^*$ moving with the velocity $v$ in relation to the proprietor B, then both the quantity of bread $A_0$, equivalent to $B_0$, and the economic distance between $A_0$ and $B_0$ will depend on this velocity in accordance with the formulas of the relativistic kinematics. In this case we can also coordinate the physical and economic spaces, if we assume that the velocity of light in the physical space equals $S_0/k$.

In the model the sense of ***limitation of the maximum velocity*** of motion of economic objects in the relativistic space of economic states becomes obvious. If, for instance, it appears that the physical and proportional to it economic distances between A and B depend on the volume of transaction (economic analog of time), so that $\left|\frac{dl_{AB}}{d \ln n_A}\right| \geq \frac{S_0}{k}$, then a contradiction arises:

- In case of increase of distance between A and B, for receiving a larger quantity of bread in point B it is more profitable to purchase in point $A$ a smaller quantity of bread manufactured there. In this case the margin of expenditures for transportation will be larger that the margin between the purchased quantities.
- In case of decrease of distance between A and B, delivery appears to be impossible at all, as the integral expenditures for the transportation require the sale of a larger quantity of bread than the transported quantity.

Such situation results in the refusal of any transportation and makes transactions impossible. In physics, a similar situation is observed for the objects located at distances exceeding the radius of the visible universe. In accordance with the Hubble formula, their relative velocity exceeds the velocity of light and exchange of any signals between them becomes impossible.

### 8.2. Model of an enclosed system of interacting companies

Previously, in paper [1], we have discussed the dynamics of manufacturing companies with rigid technological links. The main assumptions in this model included:

- The set constant coefficients $k_{ij}$ of the matrix of technologies determining the quantity of the i-th resource required for the manufacturing of the j-th product in one production cycle;
- Requirement of complete distribution of all resources manufactured in each production cycle.

In the process of fulfillment of these conditioned it turned out that both the volumes of transactions after each cycle and the corresponding relative prices (proportions of exchange) are unambiguously determined by the initial states of objects and the coefficients of the matrix of technologies. In the conditions of an extended reproduction the initial volumes of production corresponding to the proper vectors of the matrix provide their proportional exponential growth. Accordingly, the volumes of transactions, concluded upon completion of each cycle, also increase exponentially.

Thus, in the conditions of the balanced growth, the economic time determined as a logarithm of the volume of transactions increases proportionally to the physical time (quantity of the production cycles). At the same time, due to the conservation of the proportions of exchange, the economic distances between the objects remain unchanged and the system remains static in the relativistic space of economic states.

At other initial conditions, the proportions of exchange are different and become connected via the matrix of technologies with the volumes of transactions. Such connection allows associating each of the economic objects with a certain trajectory in the relativistic space of economic states. For illustration purposes, we will consider the simplest system of two companies with rigid connection described by the matrix of technologies $[k_{ij}]$. The evolution of state of each of the companies can be described as a sequence of exchanges and technological processes. As a result of exchange the qualitative state changes abruptly, while as a result of technology of production it changes smoothly. The scheme of such evolution is illustrated in Figure 12.

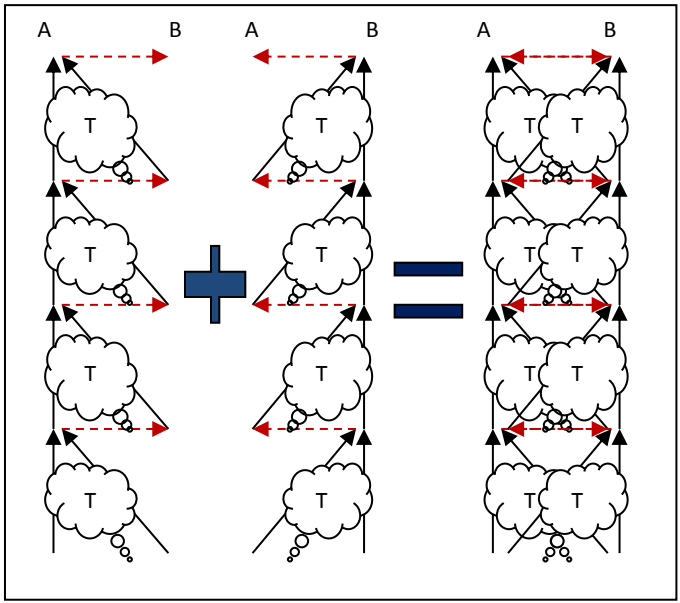

Fig.12 Scheme of interconnection of trajectories of economic objects with rigid technological connection of their manufacturers.

Let us note that in the end of each production cycle using the technology $T_A$ only a part of the product A is exchanged for the resource B, which is used in the next production cycle together with the remaining part of A. The production cycle using the technology $T_B$ is identical. As a result of superposition of these processes (their conditional summation) an analog of economic clock is originated on the basis of the repeating transaction [AB] of exchange of economic objects "A" and "B". If the initial state of the system (vector of productivities) is proportional to the proper vector of the matrix of technologies, then this "clock" will run evenly relative to the physical time, as the volumes of production and the corresponding volumes of transactions are measured in time in accordance with the exponential law. The running speed of the obtained economic "clock" is determined by the profitability, which is in this case identical for companies A and B. If the initial state of the system is not balanced, then, as it has been illustrated in [1], the evolution of such system either asymptomatically brings it to the balanced state, or results in the collapse of one of its parts (Fig.13). After a sufficient period of time, only the companies forming stable (quasi-stable) complexes of technologies will remain in the economic space. The profitability of each of them is determines the running speed of the "proper economic clock".

Due to the economic analog of the principle of relativity, any of them can be accepted as a reference point. Then the relative profitability of the rest of them will determine their relative velocity in the relativistic space of economic states in accordance with the relativistic formula of time dilation.

Thus, in the process of discussion of real economic systems, both the "economic distance" and the "economic time" can be found to be connected with the physical space and physical time. However, such connection is determined by the properties of the considered system and in each case requires additional analysis.

We can say that in real economic systems the physical time imposes a dedicated reference system, relative to which the velocities of "economic clocks" of various companies are considered. On the other hand, the physical distance inevitably influences the qualitative differences of economic objects as well. Even in case if they are otherwise indistinguishable, from the point of view of the measurement approach their economic quality is different (the same consumer will offer different price for them). Nevertheless, the economic space of "quality" is not limited only by these differences.

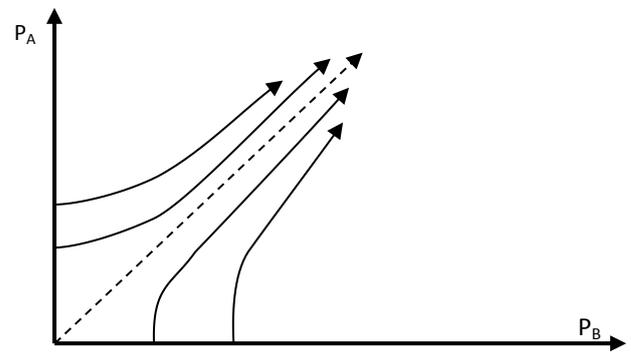

Fig. 13. Trajectories of a system consisting of two companies interconnected by rigid technological links in the coordinates of their production capacities.

## 9. Further stages of constructing the dynamics of economic objects in the framework of the measurement approach

The history of physics as an exact science is several hundred years long. At the same time, the economic theory, as a science about motion, only begins its existence. Can we expect the development of the fundamental theory of motion of economic objects to the level of physics in the near future? Undoubtedly, yes. In the first place, due to the fact that both physics and economics can be based on equivalent principles based on the properties of fundamental measurements used in these theories. Using the measurement approach we can obtain equivalent mathematical structures with similar (and often identical) laws. Thus, in the process of construction of the kinematics and dynamics of economic objects and introduction of field and quantum notions, we can use a number of ready "hints" from physics, which in due time required many years of enormous efforts of a great number of scientists.

In this chapter of the paper we will propose the economic interpretation of some of them, which seems to us rather obvious and natural. Nevertheless, let us note that the final construction of each of the chapters of the future theory of motion of economic objects is a complex task. It will require, similarly to physics, introducing the corresponding axiomatic, necessary mathematical calculations and experiments. The complexity of interpretation of the processes in the proposed economic space is conditioned by the absence of visual images. Moreover, the physical space-time imposes its images and schemes of modeling. Nevertheless, we take it upon ourselves to outline some of the trends of future researches even now.

So far we have been discussing only the kinematical tasks of the theory of pricing. They are actually limited to describing the trajectories of relative motions of some economic objects set in the same reference system relative to others. Or to calculate the trajectory using some of its generalized preset characteristics (velocity, acceleration, etc.). At the same time, we were not discussing the question of what "forces" the economic objects move along a particular trajectory in the relativistic space of economic states. Such tasks are attributed to the tasks of dynamics and require the introduction of new postulates. Above we have represented the scheme of further construction of the theory. In this paper we will limit ourselves to a brief analysis of some propositions and postulates required for this purpose.

### 9.1. Special features of using the monetary equivalent in the process of exchange and modeling of "ideal money"

The fundamental economic measurements introduced by us for modeling the results of economic measurements represent transactions of exchange of two economic objects. They completely describe the situation of natural exchange in economic relations. However, in real economics practically all such transactions are performed in two stages: exchange of product for money an exchange of money for product. Therefore, our further efforts will be concentrated on including money, as a fundamental mechanism of economic relations, into the consideration. For finding an adequate analog of the "ideal money" in the relativistic space of states, let us first analyze its economic properties.

First of all, the ideal money is not a product in the full sense of this word, as it is directly used for satisfying the needs of a subject. Any sum of ideal money earned as a result of sale has a value only insofar as it can be used for purchasing another product. As the ideal money is an abstraction, their role is always performed by real money – economic objects, the physical (and consumer) properties of which can be neglected.

Thus, the ideal money results in the notion of the "delayed transaction", previously introduced by us in the process of discussion of the fundamental economic measurements [3]. By selling a certain economic object for ideal money, the subject of the transaction acquires the economic ability of making purchases for the received sum. And vice verse, the purchaser loses this ability as he pays the ideal money for the purchased product. At the same time, the ideal money is characterized only by the quantity and does not contain information on the properties of the economic object for which it was received. Thus, the interrelations between the proprietors of various economic objects take place indirectly, by means of the ideal money.

There is an analog of such interaction between material bodies in physics – the field interaction. Instead of directly considering the forces acting between two physical bodies, an intermediary of such interaction is introduced in the form of the physical field. In case of field description it is considered that each physical body creates by its presence in a certain point of the space-time a physical field in the rest of the space, while other bodies appearing in a certain point of the space-time are influenced by the effect of this field.

Following this analogy, the ideal money in the economic space must be described not as material points (economic objects), but as fields created by them. We will state that each of the "sellers", by offering a transaction, creates a field in the surrounding space, which is unambiguously characterized by the position of the economic object offered for sale and the equivalent quantity of the ideal money. Superposition of such fields, created by the sale of particular economic objects, in each point of the space-time, unambiguously determines the possibility of purchase of the economic object located in the considered point.

Field interactions require consideration of the forces acting on the bodies, and such tasks relate to the problems of dynamics. So far the problem of possible transactions was not of interest to us. The asked "questions" ant the received "answers" were only used for determining the hierarchy of values, but did not change the properties of the economic objects or their proprietors. In case of presence of force action, we should distinguish the results of completed transactions and their influence on the economic properties of objects from the results of rejected transactions. We have previously analyzed in detail the mechanism of such influence and have shown that the rejected transactions also change the state of a subject. This influence is of information nature and should be described in the framework of the quantum-mechanical formalism [2]. Therefore, a consecutive and complete consideration of the ideal money in the relativistic space of economic states must be based on the mathematical apparatus of the quantum field theory. However, similarly to physics, in the idealized model we can neglect the influence of the rejected transactions on the economic state of the proprietors and consider a non-classical limit of such field. But even in this simplified case we need to write down the relativistic invariant equations of the classical field with account of its connection with the sources (sold economic objects). This task is beyond the framework of the first paper dedicated to the introduction of the relativistic space of economic states, and will be discussed in detail in our further publications. However, in the following chapters of the present paper we will outline the procedure of constructing the equations of dynamics in the relativistic space of economic states.

Let us note that the ideal money, which can be used for purchasing an economic object, is indistinguishable, but the decision on purchase is made by their proprietors. Therefore, the possibility of making a purchase depends not only on the summary quantity of free assets, but also on the consumer preferences of their proprietors. Below we will formalize the notion of proprietor.

### 9.2. Notion of «Proprietor» in the framework of the measurement approach

In the framework of the measurement approach we will not consider the mechanism of formation of property rights or the methods of its use. In order to introduce the category of proprietor in the discussion, it is sufficient to define him as a subject, which has rights to make decisions on a transaction of exchange of a particular economic object. This decision is made by the proprietor on the basis of a certain set of consumer preferences. We have previously associated an inertial reference system in the relativistic space of economic states with each such set.

In physics such inertial reference system can be interpreted in two ways:

- As a mathematical grid of coordinates not associated with a material object (can exist even in an empty space);
- As a physically realizable system of clocks and rules, associated with a minimally required set of material bodies.

By analogy with them we can consider two approaches to the introduction of the reference system in the relativistic space of economic states.

In the first approach the evaluation of an object does not require the ownership of this object. In the second approach a subject is able to realize his choice by exchanging the object for the ideal money or for another economic object. In this case the possibility of realization of a transaction can principally influence the subject's behavior even in case of rejection of a transaction. On the assumption of these analogies, we will define *«proprietor»* as a subject, which makes a decision on a transaction with an economic object belonging to him on the basis of a set of consumer preferences.

These preferences are set by the vector in the relativistic space of economic states.

Taking into account this quite natural definition of the notion of proprietor, the mechanism of formation of a certain set of consumer preferences of the proprietor is beyond the framework of the constructed mathematical structure. Following from the general considerations, it is obvious that it should depend on the availability of a particular property in the subject's possession. Moreover, the aforesaid generalized definition of economic objects allows attributing even the subject itself (i.e. its body, capabilities, time, etc.) to this set, at least, in the sense that any choice of a particular action of a subject can be considered as a consent or refusal of a "transaction" offered to him by the current circumstances. By accepting this hypothesis (on the possibility of modeling a proprietor using a complex system of economic objects), we can close the logical links in the chain of price formation and consider that *the set of consumer preferences is unambiguously determined by the economic properties of the available objects of property of a subject.*

At the same time, it is essential that the properties of economic objects owned by a subject are completely determined on the basis of the results of the fundamental and generalized economic measurements. Then the set of consumer preferences of a subject (proprietor) will also be completely determined by them.

### 9.3. Proceeding from the principle of least action to the principle of maximum benefit

One of the most fundamental principles of constructing the dynamics in the physical theory is the principle of least action. Considering a proper scale for a certain economic object "A", we assume that a certain consumer exists, which evaluates his subjective benefit in accordance with this scale. Keeping in mind that the construction of such scale is based on the transaction $[A_0 B_0]$, we can state that the consumer also owns a certain economic object "B", which has a secondary role for calculating the quantity of "A".

In physics a pair of mirrors can be used for a similar purpose by an observer linked with a certain material object (a spaceship, for instance). The light beam moving between them counts its own time in the reference system of the spaceship. In the economic space such scale can be the quantity of diamonds stored up by a subject in the course of his life (in carats), the number of wins in sport competitions, the number of published articles, etc. In both cases the proper scales are associated with a certain non-economic ("physical") method of measurement of the obtained benefit.

In a more general sense, each consumer can be associated with a certain technological process – a nominal "company producing subjective benefit", the owner of which is the consumer himself. The quantitative estimate of this benefit by the consumer provides the proper scale of quantity.

The reverse is also true. Each real company producing a certain product can be considered as a nominal consumer, whose set of consumer preferences is determined by the produced quantity of this product. For such company the principle of maximum benefit is limited to the production of the maximum quantity of product with the set limitations for the initial and finite state.

Summarizing this brief discussion, we can formulate the principle of maximum benefit in the following form:

- *If two states of the same economic object are set, then the transition from one state to the other is described by the dependence of quality on quantity, for which the ratio of own quantity in the finite and initial states is maximum.*

At the same time, the obtained dependence can be considered as the trajectory of motion in the relativistic space of economic states. The own quantity is estimated using a "physical" method not related to economic measurements. We will designate the logarithm of ratio of the own quantity in the finite and initial states as the increment of benefit associated with the considered economic object.

### 9.4. Two main types of interaction in physics and in economics

In the physical theory four types of fundamental interactions are known. However, in the process of consideration of the interaction of macroscopic bodies in the classical mechanics, only one of them is used – the gravitational interaction. The remaining three are latent and are observed only in the properties of solid bodies. Namely, in the forces of elasticity and friction, which are set phenomenologically. At the same time, the gravitational field is an example of the "long-range interaction", while the friction and elasticity are considered the forces of "short-range interaction". At the initial state of development of the theory of interaction in the relativistic space of economic states we will also limit ourselves only to such "mechanical" interactions. But even with such a limited description, fundamental differences between the forces of long-range interaction (gravitational forces) and short-range interaction (contact forces of elasticity and friction) are observed.

Possibly, it is a result of random coincidence, but in economics we can mark out two principally different types of interaction between economic objects as well. The first class includes the economic interactions resulting in the exchange of property as a result of transaction. We will further call them exchange interactions. The second class includes interactions not connected with the exchange of property, for instance, the influence of competition on the price formation. Such interactions are of informational nature as they only influence the evaluation by the proprietor of the economic objects in his possession, without changing their list. We can draw an analogy of such interactions with field interactions, when the influence of one body on the other is conditioned only by their presence in close proximity to each other. Moreover, we can state that the closer is the distance between these objects, the smaller are the qualitative differences between them, and the stronger is their influence on each other (they are competitors in the market).

Further development of the dynamics of economic objects in the relativistic space of economic states requires introducing new notions and studying of their properties. For instance, the economic mass, force, impulse, kinetic and potential energy, etc. We have previously passed this path in the process of consideration of the simplified (non-relativistic) economic space. Besides, in this space we used the physical time as the time axis and did not consider such notion as the qualitative differences between economic objects. Now we can review these notions in the view of the new approach to the description of the dynamics of economic objects. However, it requires a separate publication. Therefore, in this paper we will limit ourselves to the remarks on possible ways of description of these interactions and the corresponding scheme (Fig.1). As a conclusion, we will summarize the main points of constructing the kinematics in the relativistic space of economic states.

**Conclusion**

The constructed model of the relativistic space of economic states has the same advantages compared to the alternative models, as the relativistic kinematics in physics compared to common non-relativistic kinematics. Below we will list some of them. The relativistic space of economic states:

- Allows solving the main problem of price formation in systems, which do not satisfy the accepted standard assumptions, but represent economic objects in the generalized (defined above) sense.
- Provides evaluation of the "fair price" with account of the consumer's specific characteristics (subjective estimate).
- Allows rejecting the necessity of defining the system of absolute values (absolute reference system in physics) or its analog in the existing models.
- Allows including into consideration such economic objects, the properties of which cannot be described quantitatively or qualitatively using a standard approach.
- And finally, the theory of money can be constructed consistently only in the framework of the relativistic formalism, in the same way as the propagation of light requires considering the relativistic space for eliminating the paradoxes.

Generally, the relativistic space of economic states can be used as a basis for constructing the closed theory of "motion" and interaction of economic objects, i.e. the basis for modeling economic systems of any complexity.